%% file: daoinduction.tex
\pdfoutput=1
\documentclass[journal]{IEEEtran}
    
\usepackage{amsmath,epsfig,amssymb,verbatim,amsopn,cite,multirow}
\usepackage{balance}

\allowdisplaybreaks
\usepackage{tabularx}
\usepackage[usenames,dvipsnames]{color}
\usepackage{url}
\usepackage{amsfonts}
\usepackage{amssymb}
\usepackage{epsfig}
\usepackage{type1cm}
\usepackage{lettrine}
\usepackage{balance}
\usepackage[usenames,dvipsnames]{color}
\usepackage[all]{xy}  
\usepackage{url}
\usepackage{amsfonts}
\usepackage{amssymb}
\usepackage{epsfig}
\usepackage{cite}
\usepackage{graphicx}
\usepackage{epstopdf}
\usepackage{color}
\usepackage{tikz}
\usetikzlibrary{patterns}
\usepackage{array, boldline, makecell, booktabs}
\newcolumntype{?}{!{\vrule width 1pt}}
\usetikzlibrary{arrows,shapes,positioning,shadows,trees}
\newcommand{\removelatexerror}{\let\@latex@error\@gobble}
\makeatother

\include{Citations.bib}

\begin{document}
\title{The DAO Induction Attack Against the RPL-based Internet of Things}
\author{Ahmad Shabani Baghani, Sonbol Rahimpour, and Majid Khabbazian \vspace*{-0.6cm}
\thanks{The authors are with the Department of Electrical and Computer Engineering, University of Alberta, Edmonton, Canada (Email: shabanib@ualberta.ca, rahimpou@ualberta.ca, and mkhabbazian@ualberta.ca)}}
\maketitle
\begin{abstract}
RPL is the emerging routing standard for low power and lossy networks (LLNs). 
LLN is a key component of the Internet of Things (IoT),
hence its security is imperative for the age of IoT.
In this work, we present the DAO induction
attack, a novel attack against RPL.
In this attack, a malicious insider or a compromised node periodically increments its DTSN number.
Each such increment can trigger/induce a large number of control message transmissions in the network.
We show that this degrades the network performance in terms of 
end-to-end latency, packet loss ratio, and power consumption. 
To mitigate, we propose a lightweight solution to detect the DAO induction attack.
Our solution imposes nearly no overhead on IoT devices, which is important as these devices are typically constrained in terms of power, memory and processing.

\end{abstract}


\section{Introduction}

The Internet of Things (IoT) is an emerging technology which envisions to connect billions of ``things" to the Internet. This enables numerous  applications in diverse areas such as smart home, e-Health, and smart city. Low-power and Lossy Networks (LLNs) play an indispensable role in realizing IoT. These networks are typically composed of constrained devices with limited power, memory and processing. In addition, LLNs suffer from high packet loss rates and low throughput. These characteristics and limitations make designing routing protocols for LLNs challenging.   

ROLL, a working group of the Internet Engineering Task Force (IETF),  evaluated the common standard routing protocols of the Internet and concluded that these protocols are not suitable for LLNs because of their heavy overhead. Consequently, the ROLL group designed the IPv6 Routing Protocol for Low-Power and Lossy Networks (RPL)~\cite{rfc6550} to meet the low overhead requirement of LLNs. 
  
Because of limited resources, nodes in LLNs are unable to run and benefit from complex security solutions such as those that use assymetric cryptography. This makes RPL vulnerable to a range of attacks~~\cite{RaoofML19, DAOinsider , perazzo2017dio,pu2019energy, mayzaud2014study,dvir2011vera, weekly2012evaluating,wallgren2013routing,DBLP:journals/twc/KhabbazianMB09}.
An internal attacker may alter, inject, replay, and generate data or control messages to impact the normal operation of RPL networks. For example, in the version number attack~\cite{mayzaud2014study}, a malicious insider node initiates an unnecessary global network repair process by increasing the version number. Another example, albeit with lower impact, is the DAO insider attack~\cite{DAOinsider} in which the attacker repeatedly sends DAO control messages to the root, causing wasteful transmissions by the nodes on the path from the attacker to the root.

In this work, we present the DAO induction attack, a novel attack in which a malicious insider node induces nodes to transmit unnecessary DAO control messages.
Similar to the version number attack, the DAO induction attack can cause a large number of transmissions in the network. Unlike the version number attack, the DAO induction attack may not be detectable by the root of the network, as will be explained later. 

The main contributions of our work are as follows:
\begin{itemize}
    \item We introduce the DAO induction attack, a novel attack against the RPL protocol.
    \item We evaluate the impact of the DAO induction attack on power consumption, communications overhead, latency, and packet loss ratio.
    \item We propose a lightweight solution to detect the DAO induction attack.
    Our solution is fully compatible with the RPL protocol, and imposes nearly no overhead on IoT devices.
\end{itemize}

 The rest of the paper is organized as follows. Section \ref{protocol} provides an overview of the RPL routing protocol. The adversary model is given in Section~\ref{sec:Adv}. Section~\ref{DTSNattack} describes the DAO induction
attack. Section~\ref{evaluation} evaluates the attack's impact on the network performance. Section~\ref{mitigation} briefly overviews the existing mitigation techniques, and proposes a new lightweight solution to detect the DAO induction attack. Finally, Section~\ref{conclusion} concludes the paper.

\section{Overview of the RPL Protocol}
\label{protocol}
\begin{figure}
    \centering

    \includegraphics[width=0.25\textwidth]{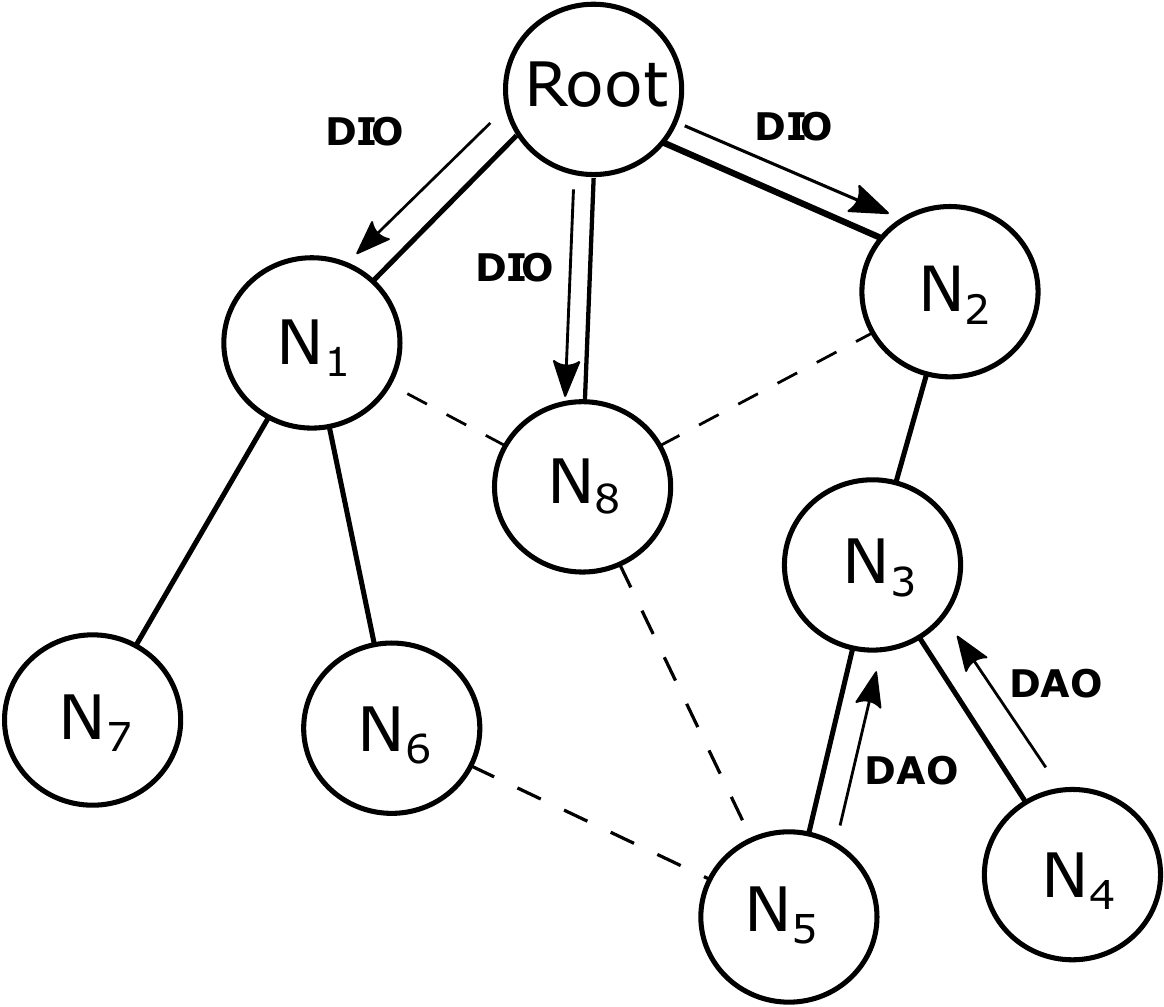}

    \caption{An example of a RPL DODAG. Solid lines show each node's preferred parents, and dashed lines show node's other DAO parents.  For instance, $\{N_3, N_6,N_8\}$ is the parent set of $N_5$, and $N_3$ is the preferred parent of $N_5$.}
    \label{RPLinstance}
\end{figure}

RPL is a distance-vector routing protocol, which can operate on various link layer standards including Bluetooth and IEEE 802.15.4~\cite{IEEE80215}. RPL builds one or more Destination Oriented Directed Acyclic Graph (DODAG),  a loop free topology as shown in Fig.~\ref{RPLinstance}. DODAG has a  single root as a destination node with no outgoing edges. The root acts as the data sink of DODAG.  Each DODAG is specified by an instance ID, a DODAG ID, and a version number. 

RPL uses DODAG to support three different traffic patterns: multipoint-to-point (MP2P) from end nodes to the root, point-to-multipoint (P2MP) from root to end nodes, and point-to-point (P2P) traffic. The DODAG structure is built step by step. To this end, nodes periodically transmit a control message called Destination Information Object (DIO).  DIO messages contain important information including an objective function to calculate rank, a number that determines  a node's position with respect to the root. 
Ranks monotonically increase in the downward direction (i.e., towards leaf nodes), and are used to avoid loops.

After receiving a DIO message from a lower rank node, the receiving node adds the address of the DIO sender to its candidate parent set, and calculates its rank with respect to the new candidate parent. The candidate parent that results in the best rank is selected as the node's preferred parent.  At the end of this procedure, each node has upward paths towards the root (through its parents). 

DIO messages are periodically sent by nodes according to a trickle timer. If a new node wants to join the network, it should receive a DIO message to obtain DODAG information. If the new node does not receive a DIO message, it can send a DODAG Information Solicitation (DIS) message requesting DODAG information. When an existing node in the network receives a DIS message, it replies by transmitting a DIO message.

To support downward routs (i.e. routs from the root), RPL uses another type of control message called Destination Advertisement Object (DAO). A node that wants to be reachable by the root advertises its address in a DAO message, and sends it to one of its DAO parents. The course of action taken by the node's DAO parent depends on the RPL mode of operation.

 RPL supports two modes of operation: storing (table-driven) and non-storing (source routing). In the storing mode, all non-leaf nodes maintain a routing table for destinations, while in the non-storing mode only the root maintains a routing table. In both modes, a node that receives a DAO message forwards it to one of its DAO parents; this ensures that the DAO message is ultimately received by the root.
In storing mode, a node updates its routing table before forwarding the DAO message. 
This update is not required in the non-storing mode as non-root nodes do not maintain any routing table.

In the non-storing mode, P2P packets travel up from the source all the way to the root and then travel down to the destination. In the storing mode, however, a P2P packet can start traveling down towards the destination as soon as it reaches a common ancestor of the source and the destination.

\section{Adversary Model}
\label{sec:Adv}
We assume that RPL is either in no-secure mode, or uses a shared secret key (at the link-layer or by itself) to secure its messages.
In either case, nodes cannot authenticate the root's messages, as every node uses the same secret key.
We assume that the adversary controls a single insider node (e.g. a compromised node), hence knows the network's secret key.
We refer to the node controlled by the adversary as the malicious node.
The malicious node can be any node in the network except the root.
In this work, we limit the malicious node's misbehaviour to 1) running the DAO induction attack (explained next), and 2) selectively dropping DAO packets to avoid detection by the root.
Attacks combining the DAO induction attack with other existing ones can be more powerful, and lie outside the scope of this work.

 \section{The DAO induction Attack}
 \label{DTSNattack}
  In RPL, each node maintains a DAO Trigger Sequence Number (DTSN), and reports it in its DIO messages.
If a node receives a DIO message from one of its DAO parents, and realizes that the parent has incremented its DTSN,
the node must schedule a DAO transmission. 
In non-storing mode, the node must in addition increment its own DTSN.
Therefore, in this mode, a DTSN increment by a node will cause all its descendants to increment their DTSN in turn, triggering DAO transmissions from the entire sub-DODAG.

In DAO induction attack, a malicious insider node repeatedly increases its DTSN to trigger DAO transmissions.
This can cause many transmissions particularly in the non-storing mode (a common mode of operation as many IoT devices
are too constrained to operate in the storing mode~\cite{rfc6550}) as all descendants of the malicious node transmit
each time the malicious node increments its DTSN.
To avoid detection by the root, the attacker can simply refrain from forwarding DAO message of its descendants to the root.


The DTSN counter is an 8-bit unsigned integer, so it has a limited range. This limitation, however, does not restrict the number of times a malicious node can update DTSN in a
DAO induction attack. This is because in RPL,  sequence counters operate according to a `lollipop' fashion~\cite{perlman1983fault}, where an increment of a sequence number with the maximum value will wrap the number back to zero.
Therefore, the number of times an attacker can increment DTSN is practically unlimited.

Similar to the version number and DAO insider attacks, the DAO induction attack can be mitigated by enabling security mechanisms
at the link-layer or at RPL itself.
However, these mechanisms are ineffective when the attacker is an insider or a compromised node.

\section{Experimental Analysis}\label{evaluation}
To evaluate the impact of the DAO induction attack on the network's performance, we performed a diverse set of simulations using the Contiki operating system~\cite{dunkels2004contiki}, a lightweight and open-source operating system designed for IoT. 
\subsection{Simulation settings}
\begin{table}[t]
\vspace{-0.2cm} 
   \caption{Simulation parameters settings}
   \label{tab:example}
   \small 
   \centering 
   \begin{tabular}{?c??c?} 
    \Xhline{0.8pt}
   \textbf{Simulation parameters} & \textbf{Value}  \\ 
   \Xhline{0.8pt} 
   Simulation time & $1800s$\\
   \Xhline{0.8pt} 
   Radio medium & Unit Disk Graph Medium \\
   \Xhline{0.8pt} 
   Topology dimension & $150m \times 150m$ \\
   \Xhline{0.8pt} 
   Number of nodes & $20$, $30$, $40$, and $50$ \\
   \Xhline{0.8pt} 
   Modes of operation & Storing and Non-storing \\
   \Xhline{0.8pt} 
   Transmission range & $40m$ \\
   \Xhline{0.8pt} 
      Interference range & $80m$\\
   \Xhline{0.8pt}
   Traffic rate per node & 1 packet per minute \\
   \Xhline{0.8pt}
   Node type & Tmote Sky\\
   \Xhline{0.8pt}
   Number of simulations & 10 per each topology \\
   \Xhline{0.8pt}
   link layer protocol & IEEE 802.15.4\\
    \Xhline{0.8pt}
     MAC protocol & CSMA-CA\\
   \Xhline{0.8pt}

   \end{tabular}
\end{table}

We used the Tmote Sky mote, which is an MSP430-based board benefiting from a radio chip compatible with the IEEE 802.15.4 link layer protocol.  We employed this mote for all the nodes including the malicious node.
To implement the DAO induction attack, we modified the RPL protocol stack of the Contiki OS on the malicious node. 
Similar to other nodes, the malicious node joins  the  network  and  actively  participates  in  the  creation  and maintenance  of  the  DODAG. The main difference between the malicious node and the others is that it is programmed to periodically increment its DTSN number, and send it in a DIO message to its neighbours.
To evaluate the maximum impact of the DAO induction attack in the non-storing mode, we selected the malicious node randomly from the neighbours of the root.
Note that these nodes have the maximum number of descendants among all non-root nodes.

We considered a sample scenario in which nodes are distributed randomly in a $150m \times 150m$ square area network.    
Each node is static and transmits one data packet of $50$ bytes to the root every  $60$ seconds. 
To simulate link failure, we used the Unit DISK Graph Model (UDGM).
We evaluated the impacts of the DAO attack on the following metrics.

\begin{itemize}
\item \textit{DAO overhead:} the total number of DAO transmissions including transmissions of original DAO messages as well as transmissions for forwarding DAO messages towards the root.
\item \textit{Average power consumption:} the average consumed power by each node in the network.
\item \textit{Packet loss ratio:} the packet loss ratio averaged over all the node in the network.   The packet loss ratio of a node is one minus the ratio of the number of received packets by the DODAG root from that node over the total number of packets sent by the node.  
\item \textit{Average Latency:} the average end-to-end latency
of all packets successfully received by the root. 
\end{itemize}

\subsection{Impact of the DAO induction attack}
 \begin{figure}[t]
    \centering
    \setlength\abovecaptionskip{-0.3\baselineskip}
    \includegraphics[width=0.43\textwidth]{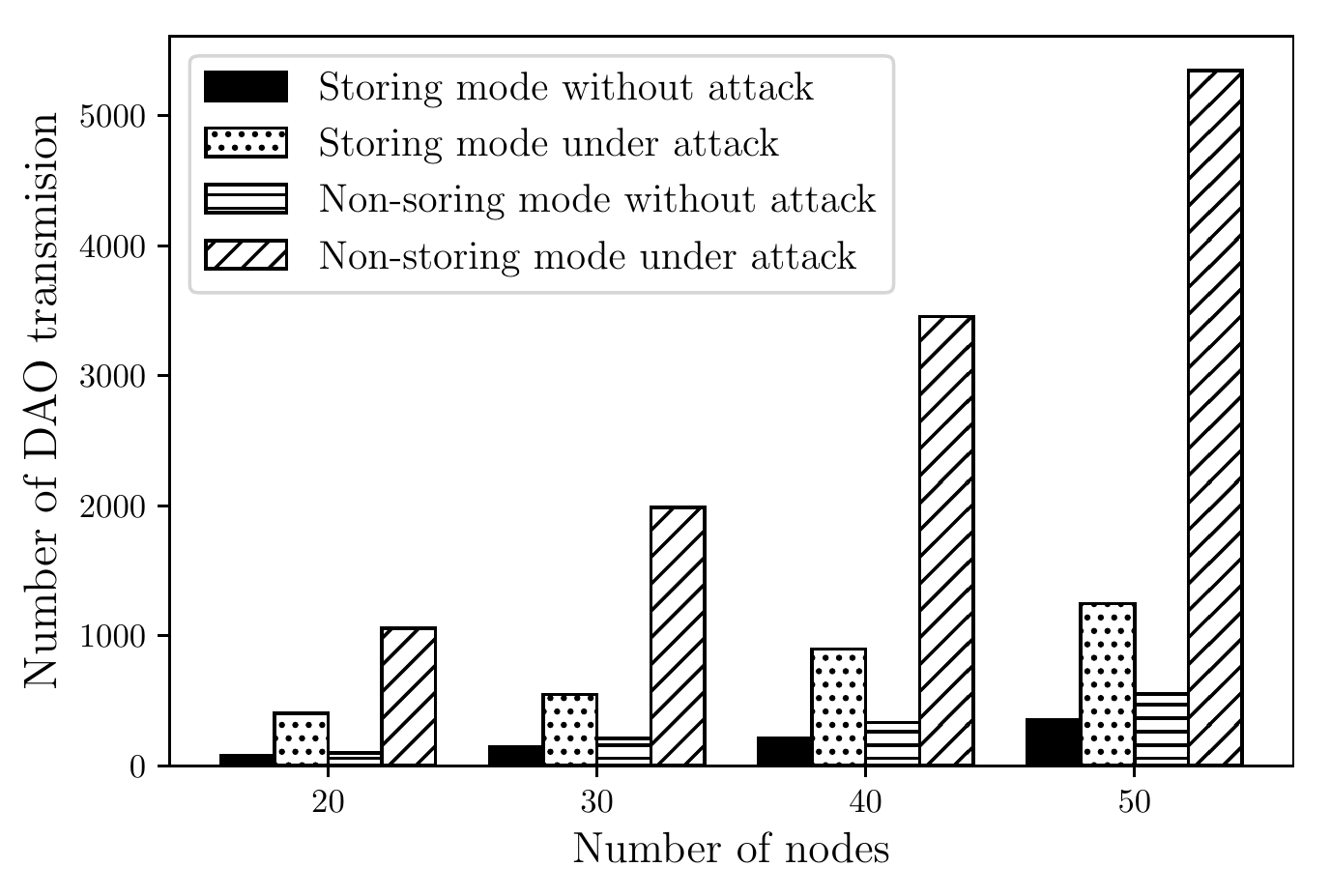}
    \caption{The impact of the DAO induction attack on the number of DAO transmissions in the storing and non-storing modes.}
    \label{dao}
\end{figure}

\begin{figure}[t]
    \centering
    \setlength\abovecaptionskip{-0.3\baselineskip}
    \includegraphics[width=0.43\textwidth]{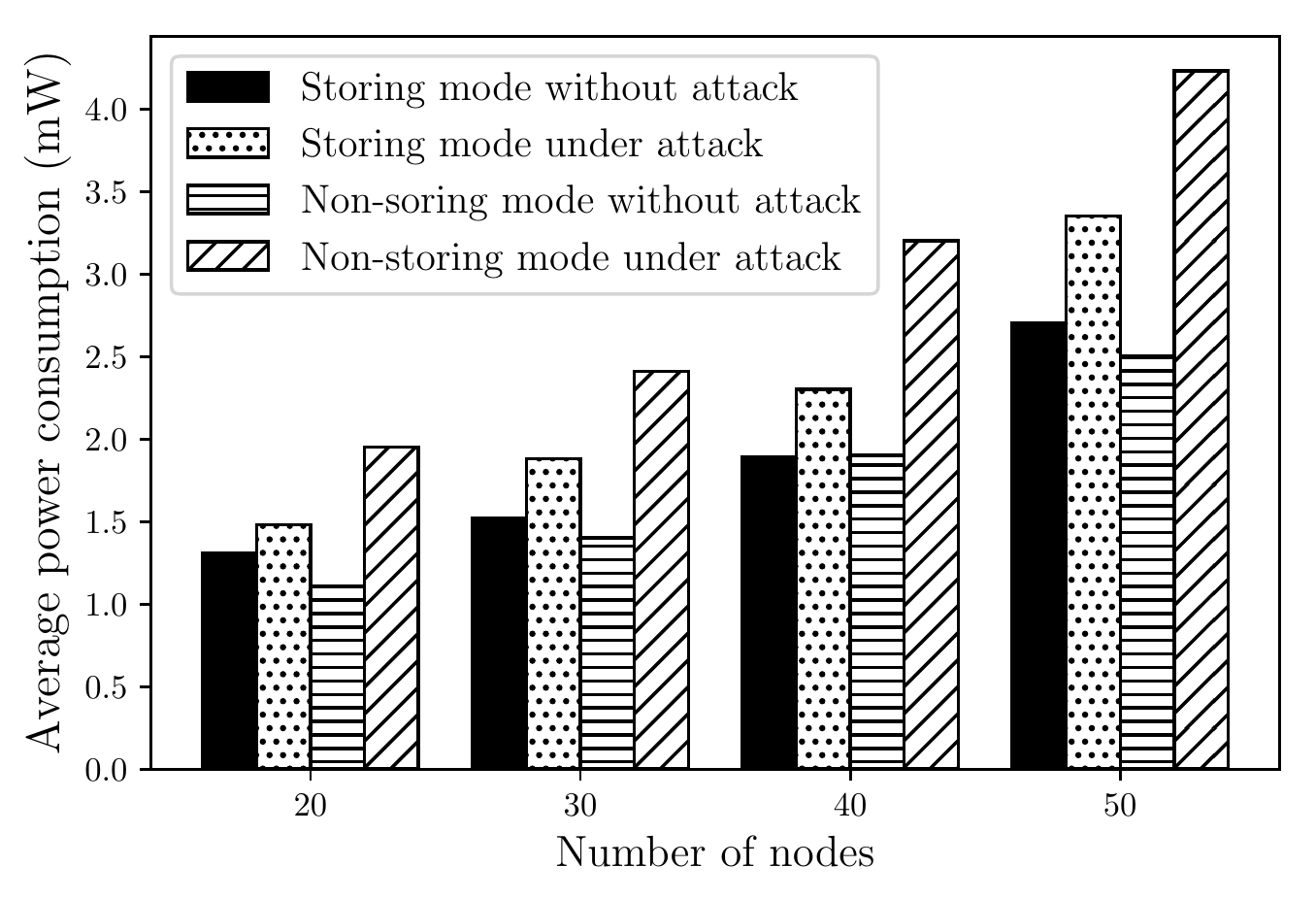}
    \caption{The impact of the DAO induction attack on the average power consumption in the storing and non-storing modes.}
    \label{power}
\end{figure}

\begin{figure}[t]
    \centering
    \setlength\abovecaptionskip{-0.3\baselineskip}
    \vspace{0.1cm} 
    \includegraphics[width=0.43\textwidth]{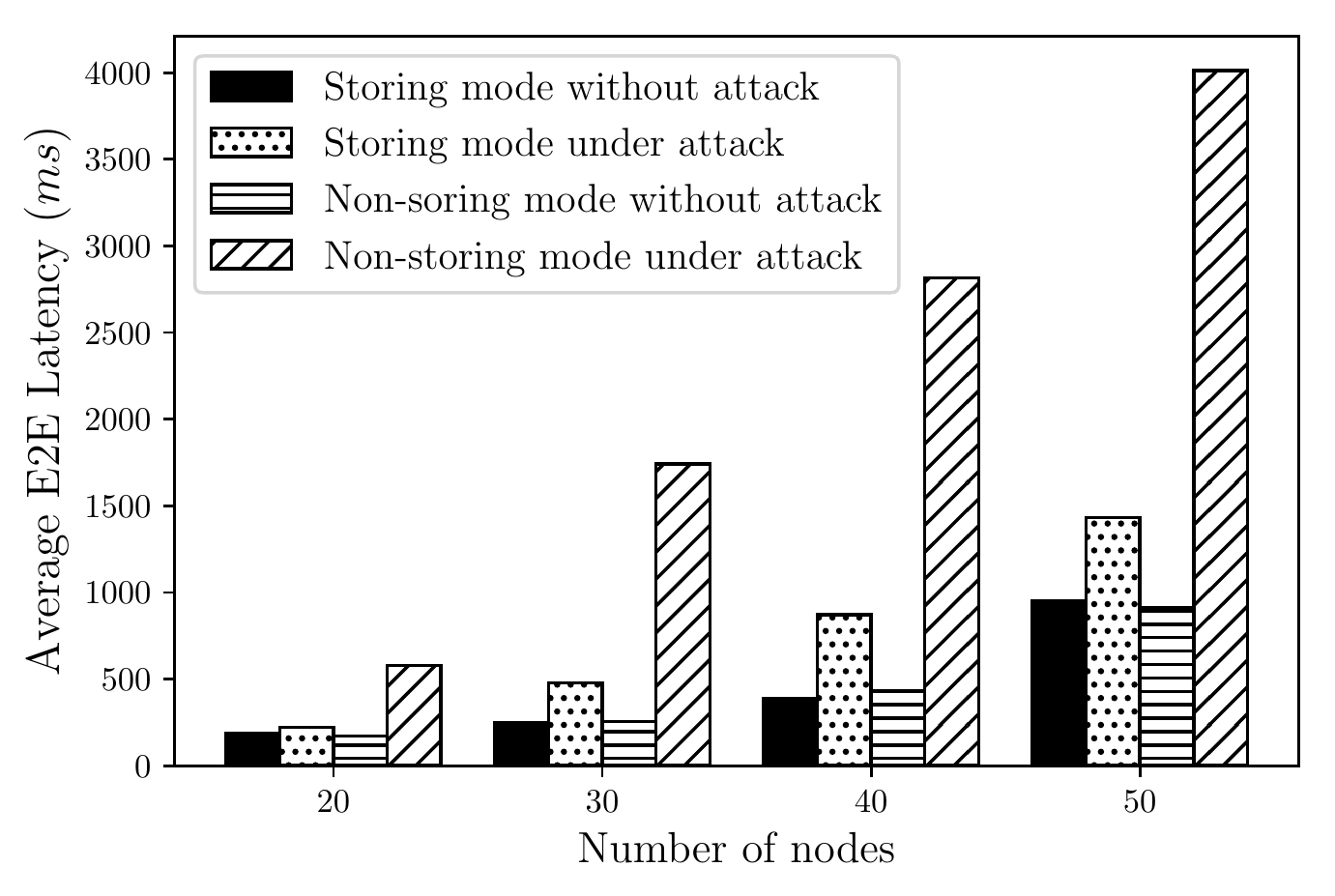}
    \caption{The impact of the DAO induction attack on the average end-to-end latency in the storing and non-storing modes.}
    \label{latency}
\end{figure}

\begin{figure}[t]
    \centering
    \setlength\abovecaptionskip{-0.3\baselineskip}
    \includegraphics[width=0.43\textwidth]{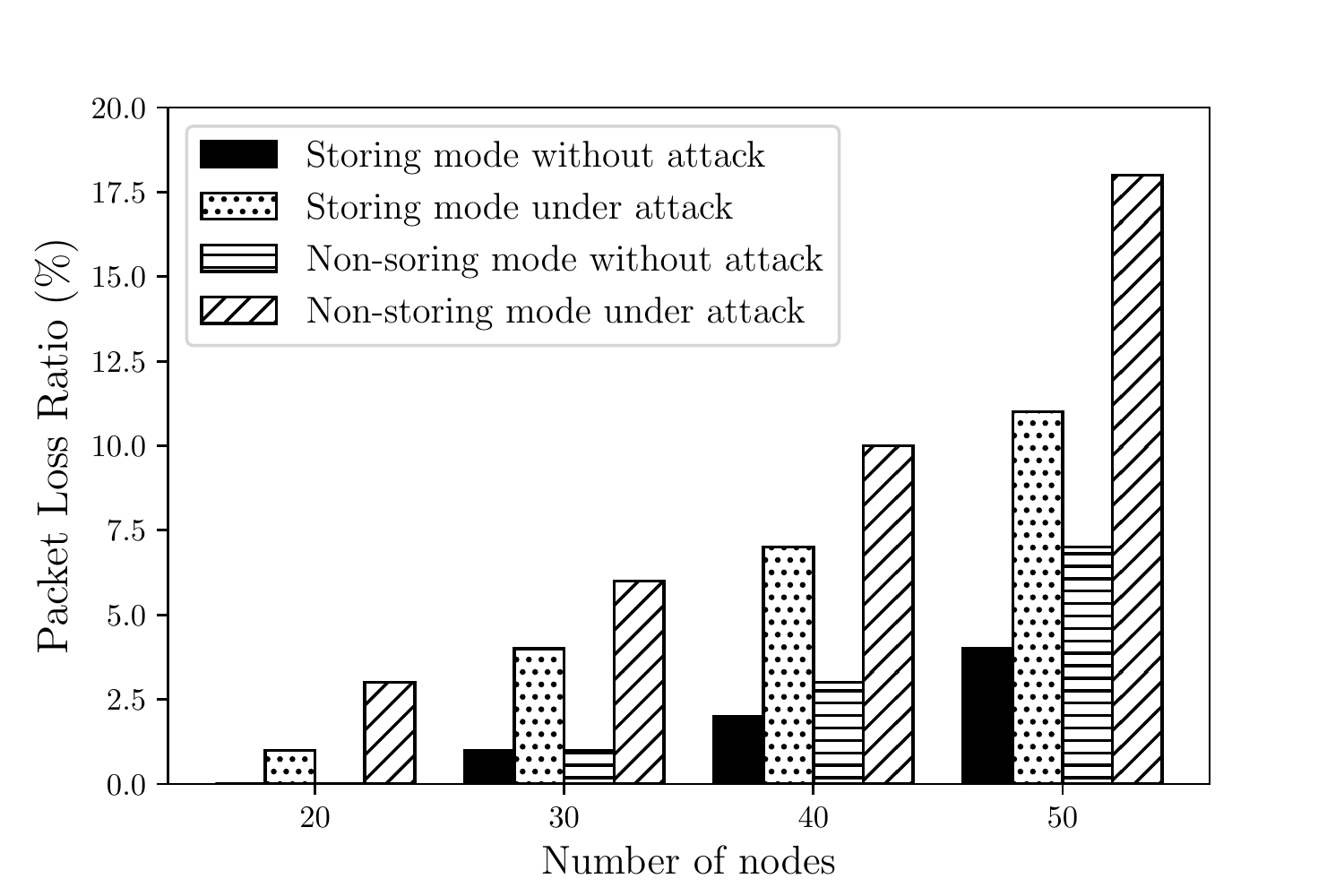}
    \caption{The impact of the DAO induction attack on the packet loss ratio in the storing and non-storing modes.}
    \label{PLR}
\end{figure}
 Fig.~\ref{dao} shows the total number of DAO transmissions (i.e., the DAO overhead) for both RPL modes of operation.  As shown, the DAO induction attack significantly increases the DAO overhead in both storing and non-storing modes. In larger networks, this overhead is higher: When there is no attack, the DAO overhead increases slowly with the number of nodes. Under the DAO induction attack, however, the DAO overhead grows at a significantly higher rate.  Note that, the impact of the DAO induction attack is higher in the non-storing mode than the storing mode. This is expected because, in the non-storing mode, a DTSN increment triggers all the nodes in the attacker's sub-DODAG to transmit DAO messages.
 
 Fig.~\ref{power} shows the average power consumption of nodes when the network is under the DAO induction attack.  
 To calculate the average power consumption, we used the collect-view feature 
available in Contiki. As shown, the power consumption increase because of the DAO induction attack is more noticeable in the non-storing mode than in the storing mode.  This is expected because, in the non-storing,  the attack engages more nodes and generates more overhead.  

Fig.~\ref{latency} shows the impact of the DAO induction attack on the average end-to-end latency. As shown in the figure, the DAO induction attack significantly increases the average end-to-end delay in the network. 
This increase is considerably higher in the non-storing mode than the storing mode. Again, the underlying reason is that the DAO induction attack engages more nodes and creates more overhead in the non-storing mode.

Finally, the impact of the DAO attack on the packet loss ratio is shown in Fig.~\ref{PLR}. This impact is insignificant in small networks particularly in the storing mode. The impact is, however, considerable in networks with about 40 and more nodes.  As in the previous cases,  the DAO induction attack is more severe in the non-storing mode than in the storing mode.

\section{Mitigation }\label{mitigation}
Following, we first enumerate the existing mitigation solutions by categorizing them into two classes: proactive and reactive.
We then present our solution to detect the DAO induction attack. This solution, unlike the existing ones, imposes nearly no overhead on IoT devices, which is important as these devices typically have limited resources. 

\subsection{Proactive solutions}
Proactive solutions aim to eliminate the possibility of the attack completely.
Recall that the impact of the DAO induction attack is significantly more severe in the non-storing mode than the storing mode. Theretofore, it is more important to mitigate this attack in the non-storing mode.

In the non-storing mode, the DAO induction attack is similar to the version number attack as both the version number and DTSN must be first incremented by the root. Hence both attacks can be prevented if root's messages can be authenticated.
Authentication can be achieved using digital signatures, or hash chains as described below. 

\subsubsection{Digital signatures}
the conventional way to provide authentication is by digital signatures. Use of digital signatures in IoT networks is challenging. First challenge is to securely distribute the root's public key. Currently, manual installation is the only feasible method to distribute security keys among constrained devices~\cite{raza2016s3k}. Another major challenge is that existing digital signature methods are computationally heavy~\cite{mossinger2016towards}. 

\subsubsection{Hash-chain}
 as used in VeRA~\cite{dvir2011vera}, hash chains can be used for authentication. Similar to digital signatures, hash chain based solutions impose communication and computation overheads even in normal conditions when network is under no attack. 
More importantly, for this solutions to work, the root of the hash chain must be securely distributed.
In the absence of computationally heavy asymmetric cryptography operations -- as constrained nodes have difficulty performing these operations -- the daunting manual installation seems to be the only feasible option.

\subsection{Reactive solutions}
Reactive solutions, unlike proactive ones, do not eliminate the possibility of the attack. Instead, they aim to detect and mitigate the attack upon detection. A reactive security solution consists of two phases: detection and reaction. The aim of the first phase is to detect the onset of the attack
by monitoring the network. When an attack is detected, the solution goes to the reaction phase where the attacking node is isolated/removed. 

Monitoring of the network can be performed by either the internal IoT nodes, or external monitoring nodes. 
Each of these approaches have their own issues.
The former approach imposes overheads on IoT devices, which is not desirable if they are power constrained (e.g., when they run on batteries).
The latter approach can be costly particularly when multiple external monitoring nodes are needed. 

Our proposed monitoring solution, presented next, uses the root node for detecting the DAO attack, and imposes nearly no overhead on IoT devices. In addition, simulation results show that our solution has a high detection rate.

\subsection{Our proposed detection solution}\label{proposal}
As mentioned earlier, the existing solutions all impose overhead even in normal condition when the network is under no attack. Our detection solution, however, imposes nearly no overhead. 

Our solution requires IoT nodes to follow two simple rules, both supported by the RPL protocol. First, each node should select up to two non-preferred parent nodes, whenever such nodes exist.
Second, each node should schedule its DAO transmission to be forwarded to its preferred DAO parent when it hears a DTSN increment by a non-preferred parent.  
\begin{figure}[t]
    \centering
    \includegraphics[width=0.35\textwidth]{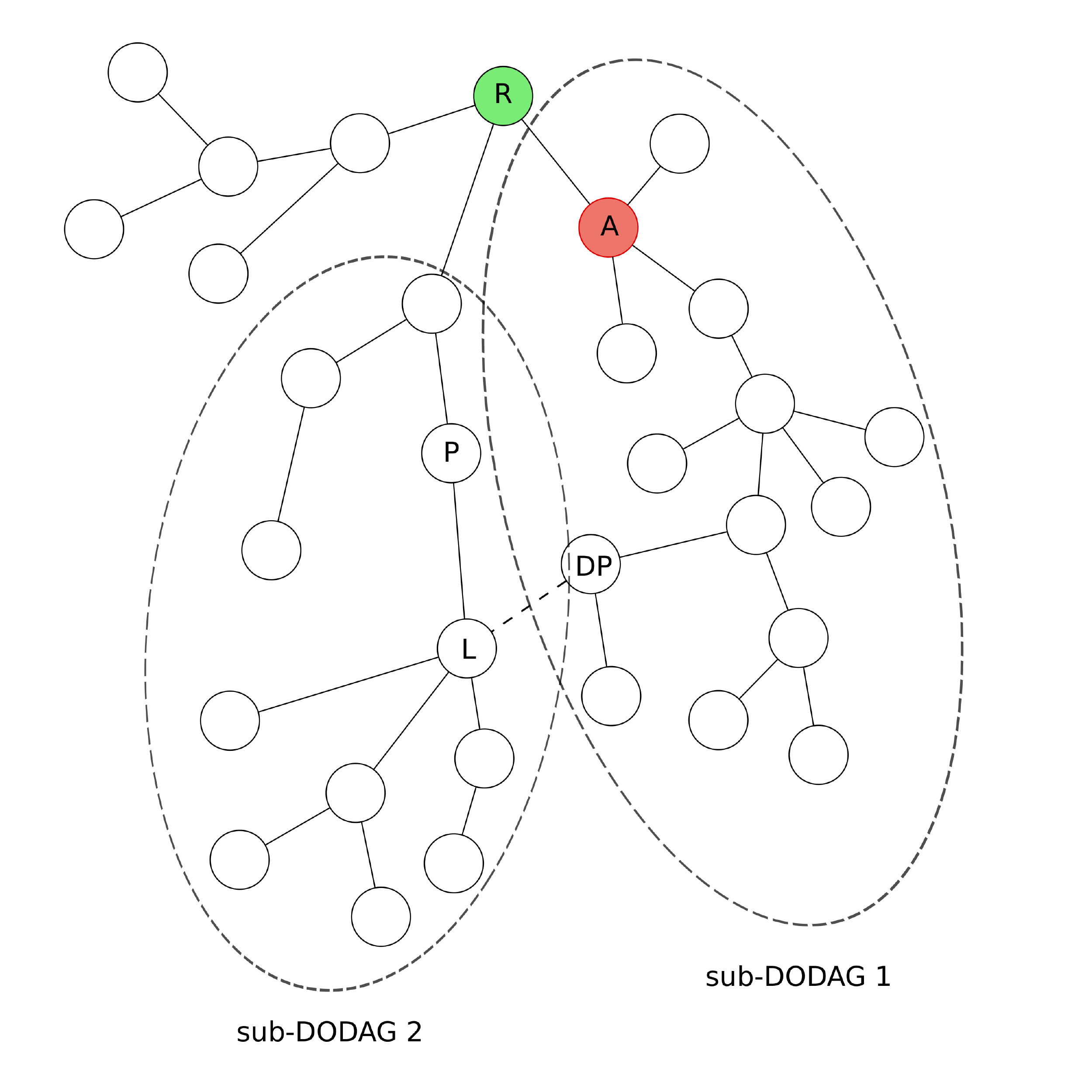}
    \caption{An example of a DODAG under the DAO induction attack by node $A$. Node ``P'' is the preferred parent and node ``DP'' is the DAO parent of node ``L'' respectively.}
    \label{leakage}
\end{figure}

Let us use an example to explain why following these rules helps the root to detect the DAO induction attack.  Consider the sample network shown in Fig.~\ref{leakage}. Node $A$ is the attacker, and the network operates in the non-storing mode. Notice that node $L$ has two DAO parents: the preferred parent $P$, and the non-preferred parent $DP$.
When the attacker  $A$ increments its DTSN, all its descendants including node $DP$ increment their DTSN in turn, and report this change through DIO messages. When node $L$ hears $DP$'s DIO message, it schedules a DAO transmissions through its preferred parent $P$ instead of $DP$. This DAO message cannot be dropped by the attacker as it does not go through the attacker.  
The root will then receive the DAO message and detect the DAO induction attack as it did not start the DTSN 
increment\footnote{Note that a single bit in DAO message can indicate that the message was generated as the result of a DTSN increment.}.
Note that when the DTSN increment is legitimate (i.e., it is started by the root), all the nodes in the network will schedule a DAO transmission.
Therefore, following the aforementioned rules does not impose any extra overhead on IoT devices.

To detect the DAO induction attack, the network should have a node (like $L$)
with two DAO parrents; one in the attacker's sub-DODAG, and the preferred one outside the attacker's sub-DODAG.
An interesting question is how often such a node exists.
To answer this question, we run simulations using the Contiki operating system.
We changed the RPL setting of Contiki, to allow nodes select more than one DAO parent whenever possible. 

Let $u_1, u_2, \ldots, u_n$ be the set of non-root IoT nodes,
and $d_i$, $1\leq i \leq n$, be a binary number, such that it is equal to 1 iff a DAO induction attack by node $u_i$ is detectable.
The detection rate is calculated as the weighted average of $d_i$, where weight of $d_i$ is the number of descendants of $u_i$ (i.e., the number of nodes that are affected by the DAO attack launched by $u_i$).
\begin{figure}[t]
    \centering
    \setlength\abovecaptionskip{-0.3\baselineskip}
    \includegraphics[width=0.43\textwidth]{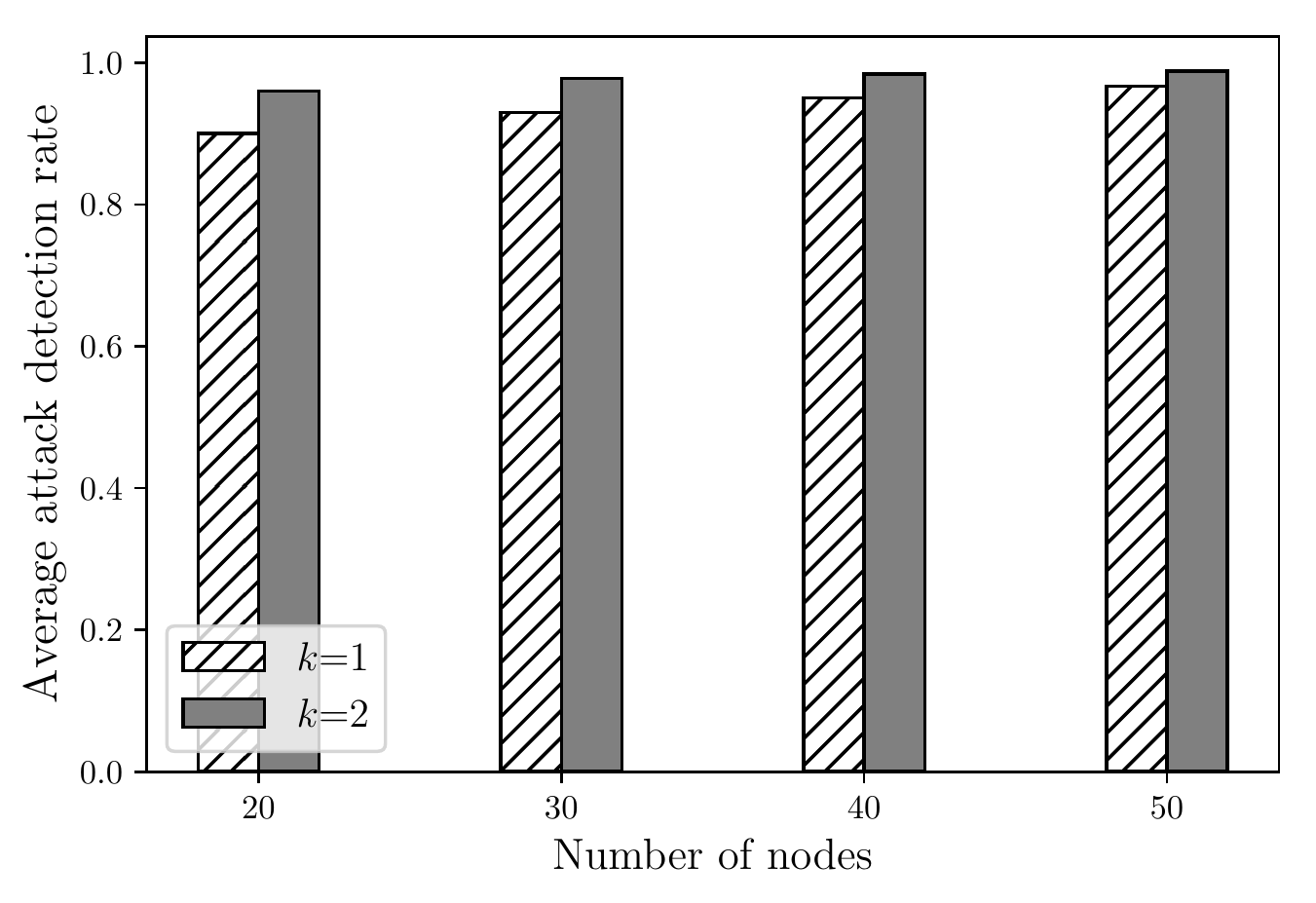}
    \caption{The detection ratio of the DAO induction attack when RPL nodes have more than one DAO parent. Each RPL node are able to choose $k$ extra DAO parents. }
    \label{detection}
\end{figure}
For a given network size, the detection rate is calculated for ten different networks of that size. 
Fig.~\ref{detection}. shows the average of these ten detection rates for network size of 20, 30, 40 and 50. In the figure, $k$ indicates the number of non-preferred DAO parents that each RPL node can have. For instances, for $k=1$, each node selects
exactly one non-preferred DAO parent whenever possible. 
As shown, the detection rate of our solution is close to $100\%$ if nodes select two non-preferred DAO parents whenever possible.

\section{Conclusion}\label{conclusion}
In this paper, we introduced the DAO induction attack, a novel security attack against the RPL protocol in which a malicious insider node increments its DTSN number periodically to flood the network with control messages.  Through various simulations, we showed that the attack adversely impacts network performance, and power consumption particularly when the network operates in the non-storing mode. To mitigate, we proposed a lightweight detection solution that imposes nearly no overhead on IoT devices. Simulation results show that our solution can detect the DAO induction attack with high probability.

\bibliographystyle{ieeetr}
\bibliography{Citations}

\end{document}

%% file: Citations.bib
@inproceedings{chugh2012case,
  title={Case study of a black hole attack on LoWPAN-RPL},
  author={Chugh, Karishma and Aboubaker, L and Loo, Jonathan},
  booktitle={Proc. of the Sixth International Conference on Emerging Security Information, Systems and Technologies (SECURWARE), Rome, Italy},
  pages={157--162},
  year={2012}
}
@article{IEEE80215,
  title={{IEEE} std. 802.15.4: Wireless Medium Access Control  and Physical Layer 
Specifications for Low Rate Wireless Personal Area Networks (LR-WPANs)},
  author={IEEE},
  year={2015},
  publisher={IEEE}
}
@inproceedings{bauer2016ecdsa,
  title={ECDSA on things: IoT integrity protection in practise},
  author={Bauer, Johannes and Staudemeyer, Ralf C and P{\"o}hls, Henrich C and Fragkiadakis, Alexandros},
  booktitle={International Conference on Information and Communications Security},
  pages={3--17},
  year={2016},
  organization={Springer}
}
@techreport{ietf-roll-rpl-observations-03,
	number =	{draft-ietf-roll-rpl-observations-03},
	type =		{Internet-Draft},
	institution =	{Internet Engineering Task Force},
	publisher =	{Internet Engineering Task Force},
	note =		{Work in Progress},
	url =		{https://datatracker.ietf.org/doc/html/draft-ietf-roll-rpl-observations-03},
        author =	{RAHUL ARVIND JADHAV and Rabi Narayan Sahoo and Yuefeng Wu},
	title =		{{RPL Observations}},
	pagetotal =	18,
	year =		2019,
	month =		nov,
	day =		26,
	abstract =	{This document describes RPL protocol design issues, various observations and possible consequences of the design and implementation choices.},
}
@article{perlman1983fault,
  title={Fault-tolerant broadcast of routing information},
  author={Perlman, Radia},
  journal={Computer Networks (1976)},
  volume={7},
  number={6},
  pages={395--405},
  year={1983},
  publisher={Elsevier}
}
@inproceedings{mossinger2016towards,
  title={Towards quantifying the cost of a secure {IoT}: Overhead and energy consumption of {ECC} signatures on an {ARM}-based device},
  author={M{\"o}ssinger, Max and Petschkuhn, Benedikt and Bauer, Johannes and Staudemeyer, Ralf C and W{\'o}jcik, Marcin and P{\"o}hls, Henrich C},
  booktitle={IEEE 17th International Symposium on A World of Wireless, Mobile and Multimedia Networks (WoWMoM)},
  pages={1--6},
  year={2016},
  organization={}
}

@article{le2016specification,
  title={A specification-based IDS for detecting attacks on RPL-based network topology},
  author={Le, Anhtuan and Loo, Jonathan and Chai, Kok and Aiash, Mahdi},
  journal={Information},
  volume={7},
  number={2},
  pages={25},
  year={2016},
  publisher={Multidisciplinary Digital Publishing Institute}
}
@article{raoof2019secure,
  title={Secure Routing in IoT: Evaluation of RPL Secure Mode under Attacks},
  author={Raoof, Ahmed and Matrawy, Ashraf and Lung, Chung-Horng},
  journal={arXiv preprint arXiv:1905.10314},
  year={2019}
}
@article{kamgueu2018survey,
  title={Survey on RPL enhancements: a focus on topology, security and mobility},
  author={Kamgueu, Patrick Olivier and Nataf, Emmanuel and Ndie, Thomas Djotio},
  journal={Computer Communications},
  volume={120},
  pages={10--21},
  year={2018},
  publisher={Elsevier}
}
@article{levis2011trickle,
  title={The trickle algorithm},
  author={Levis, Philip and Clausen, Thomas Heide},
  year={2011}
}
@incollection{el2016performance,
  title={On the performance of key pre-distribution for rpl-based iot networks},
  author={El Hajjar, Ayman and Roussos, George and Paterson, Maura},
  booktitle={Interoperability, Safety and Security in IoT},
  pages={67--78},
  year={2016},
  publisher={Springer}
}

@article{diffie1976new,
  title={New directions in cryptography},
  author={Diffie, Whitfield and Hellman, Martin},
  journal={IEEE transactions on Information Theory},
  volume={22},
  number={6},
  pages={644--654},
  year={1976},
  publisher={IEEE}
}
@inproceedings{raza2011securing,
  title={Securing communication in 6LoWPAN with compressed IPsec},
  author={Raza, Shahid and Duquennoy, Simon and Chung, Tony and Yazar, Dogan and Voigt, Thiemo and Roedig, Utz},
  booktitle={2011 International Conference on Distributed Computing in Sensor Systems and Workshops (DCOSS)},
  pages={1--8},
  year={2011},
  organization={IEEE}
}
@article{arics2019new,
  title={New lightweight mitigation techniques for RPL version number attacks},
  author={Ar{\i}{\c{s}}, Ahmet and Yal{\c{c}}{\i}n, S{\i}dd{\i}ka Berna {\"O}rs and Oktu{\u{g}}, Sema F},
  journal={Ad Hoc Networks},
  volume={85},
  pages={81--91},
  year={2019},
  publisher={Elsevier}
}
@inproceedings{eschenauer2002key,
  title={A key-management scheme for distributed sensor networks},
  author={Eschenauer, Laurent and Gligor, Virgil D},
  booktitle={Proceedings of the 9th ACM conference on Computer and communications security},
  pages={41--47},
  year={2002},
  organization={ACM}
}

@article{ishaq2013ietf,
  title={IETF standardization in the field of the internet of things (IoT): a survey},
  author={Ishaq, Isam and Carels, David and Teklemariam, Girum K and Hoebeke, Jeroen and Abeele, Floris Van den and Poorter, Eli De and Moerman, Ingrid and Demeester, Piet},
  journal={Journal of Sensor and Actuator Networks},
  volume={2},
  number={2},
  pages={235--287},
  year={2013},
  publisher={Multidisciplinary Digital Publishing Institute}
}
@inproceedings{da2005decentralized,
  title={Decentralized intrusion detection in wireless sensor networks},
  author={da Silva, Ana Paula R and Martins, Marcelo HT and Rocha, Bruno PS and Loureiro, Antonio AF and Ruiz, Linnyer B and Wong, Hao Chi},
  booktitle={Proceedings of the 1st ACM international workshop on Quality of service \& security in wireless and mobile networks},
  pages={16--23},
  year={2005}
}
@article{perrey2013trail,
  title={{TRAIL}: Topology authentication in {RPL}},
  author={Perrey, Heiner and Landsmann, Martin and Ugus, Osman and Schmidt, Thomas C and W{\"a}hlisch, Matthias},
  journal={arXiv preprint arXiv:1312.0984},
  year={2013}
}

@article{le20126lowpan,
  title={{6LoWPAN}: a study on QoS security threats and countermeasures using intrusion detection system approach},
  author={Le, Anhtuan and Loo, Jonathan and Lasebae, Aboubaker and Aiash, Mahdi and Luo, Yuan},
  journal={International Journal of Communication Systems},
  volume={25},
  number={9},
  pages={1189--1212},
  year={2012},
  publisher={Wiley Online Library}
}
@inproceedings{kasinathan2013denial,
  title={Denial-of-Service detection in 6LoWPAN based Internet of Things},
  author={Kasinathan, Prabhakaran and Pastrone, Claudio and Spirito, Maurizio A and Vinkovits, Mark},
  booktitle={Wireless and Mobile Computing, Networking and Communications (WiMob), 2013 IEEE 9th International Conference on},
  pages={600--607},
  year={2013}
}
@inproceedings{kim2008protection,
  title={Protection against packet fragmentation attacks at 6lowpan adaptation layer},
  author={Kim, HyunGon},
  booktitle={Convergence and Hybrid Information Technology, 2008. ICHIT'08. International Conference on},
  pages={796--801},
  year={2008}
}
@inproceedings{raza2012lightweight,
  title={Lightweight IKEv2: a key management solution for both the compressed IPsec and the IEEE 802.15. 4 security},
  author={Raza, Shahid and Voigt, Thiemo and Jutvik, Vilhelm},
  booktitle={Proceedings of the IETF workshop on smart object security},
  volume={23},
  year={2012}
}
@article{brandt2016applicability,
  title={Applicability statement: The use of the routing protocol for low-power and lossy networks ({RPL}) protocol suite in home automation and building control},
  author={Brandt, A and Baccelli, E and Cragie, R and van der Stok, P},
  journal={Consultant},
  year={2016}
}
@article{kent2005ip,
  title={IP authentication header},
  author={Kent, Stephen},
  year={2005}
}
@techreport{loughney2006ipv6,
  title={Ipv6 node requirements},
  author={Jankiewicz, Edward and Loughney, John and Narten, Thomas},
  year={2011}
}
@inproceedings{granjal2010secure,
  title={A secure interconnection model for IPv6 enabled wireless sensor networks},
  author={Granjal, Jorge and Monteiro, Edmundo and Silva, Jorge S{\'a}},
  booktitle={Wireless Days (WD), 2010 IFIP},
  pages={1--6},
  year={2010}
}
@article{mayzaud2017distributed,
  title={A distributed monitoring strategy for detecting version number attacks in {RPL-based} networks},
  author={Mayzaud, Anth{\'e}a and Badonnel, R{\'e}mi and Chrisment, Isabelle},
  journal={IEEE Transactions on Network and Service Management},
  volume={14},
  number={2},
  pages={472--486},
  year={2017},
  publisher={IEEE}
}

@article{raza2014secure,
  title={Secure communication for the Internet of Things—a comparison of link-layer security and IPsec for 6LoWPAN},
  author={Raza, Shahid and Duquennoy, Simon and H{\"o}glund, Joel and Roedig, Utz and Voigt, Thiemo},
  journal={Security and Communication Networks},
  volume={7},
  number={12},
  pages={2654--2668},
  year={2014},
  publisher={Wiley Online Library}
}
@article{raza2018tinyike,
  title={{TinyIKE}: Lightweight {IKEv2} for Internet of Things},
  author={Raza, Shahid and Magn{\'u}sson, R{\'u}nar M{\'a}r},
  journal={IEEE Internet of Things Journal},
  volume={6},
  number={1},
  pages={856--866},
  year={2018},
  publisher={IEEE}
}
@article{raza2016s3k,
  title={S3K: Scalable security with symmetric keys—{DTLS} key establishment for the Internet of Things},
  author={Raza, Shahid and Seitz, Ludwig and Sitenkov, Denis and Selander, G{\"o}ran},
  journal={IEEE Transactions on Automation Science and Engineering},
  volume={13},
  number={3},
  pages={1270--1280},
  year={2016},
  publisher={IEEE}
}
@article{yang2017survey,
  title={A survey on security and privacy issues in internet-of-things},
  author={Yang, Yuchen and Wu, Longfei and Yin, Guisheng and Li, Lijie and Zhao, Hongbin},
  journal={IEEE Internet of Things Journal},
  volume={4},
  number={5},
  pages={1250--1258},
  year={2017},
  publisher={IEEE}
}
@inproceedings{ronen2016extended,
  title={Extended functionality attacks on iot devices: The case of smart lights},
  author={Ronen, Eyal and Shamir, Adi},
  booktitle={Security and Privacy (EuroS\&P), 2016 IEEE European Symposium on},
  pages={3--12},
  year={2016}
}
@article{lerankattack,
  title={The impact of rank attack on network topology of routing protocol for low-power and lossy networks},
  author={Le, Anhtuan and Loo, Jonathan and Lasebae, Aboubaker and Vinel, Alexey and Chen, Yue and Chai, Michael},
  journal={IEEE Sensors Journal},
  volume={13},
  number={10},
  pages={3685--3692},
  year={2013},
  publisher={IEEE}
}
@inproceedings{andrea2015internet,
  title={Internet of Things: Security vulnerabilities and challenges},
  author={Andrea, Ioannis and Chrysostomou, Chrysostomos and Hadjichristofi, George},
  booktitle={Computers and Communication (ISCC), 2015 IEEE Symposium on},
  pages={180--187},
  year={2015}
}
@article{abomhara2015cyber,
  title={Cyber security and the internet of things: vulnerabilities, threats, intruders and attacks},
  author={Abomhara, Mohamed and others},
  journal={Journal of Cyber Security and Mobility},
  volume={4},
  number={1},
  pages={65--88},
  year={2015},
  publisher={River Publishers}
}
@inproceedings{cho2009attack,
  title={Attack model and detection scheme for Botnet on 6LoWPAN},
  author={Cho, Eung Jun and Kim, Jin Ho and Hong, Choong Seon},
  booktitle={Asia-Pacific Network Operations and Management Symposium},
  pages={515--518},
  year={2009}
}
@article{rghioui2014denial,
  title={Denial-of-Service attacks on 6LoWPAN-RPL networks: Threats and an intrusion detection system proposition},
  author={Rghioui, Anass and Khannous, Anass and Bouhorma, Mohammed},
  journal={Journal of Advanced Computer Science \& Technology},
  volume={3},
  number={2},
  pages={143},
  year={2014},
  publisher={Science Publishing Corporation}
}
@inproceedings{rghioui2013analytical,
  title={Analytical study of security aspects in 6LoWPAN networks},
  author={Rghioui, Anass and Bouhorma, Mohammed and Benslimane, Abderrahim},
  booktitle={Information and Communication Technology for the Muslim World (ICT4M), 5th International Conference on},
  pages={1--5},
  year={2013}
}
@article{zhang2014sybil,
  title={Sybil attacks and their defenses in the internet of things},
  author={Zhang, Kuan and Liang, Xiaohui and Lu, Rongxing and Shen, Xuemin},
  journal={IEEE Internet of Things Journal},
  volume={1},
  number={5},
  pages={372--383},
  year={2014},
  publisher={IEEE}
}
@inproceedings{kamble2017security,
  title={Security attacks and secure routing protocols in RPL-based Internet of Things: Survey},
  author={Kamble, Arvind and Malemath, Virendra S and Patil, Deepika},
  booktitle={Emerging Trends \& Innovation in ICT (ICEI), 2017 International Conference on},
  pages={33--39},
  year={2017}
}
@inproceedings{xie2010routing,
  title={Routing loops in dag-based low power and lossy networks},
  author={Xie, Weigao and Goyal, Mukul and Hosseini, Hossein and Martocci, Jerald and Bashir, Yusuf and Baccelli, Emmanuel and Durresi, Arjan},
  booktitle={Advanced Information Networking and Applications (AINA), 2010 24th IEEE International Conference on},
  pages={888--895},
  year={2010}
}
@inproceedings{mayzaud2014study,
  title={A study of {RPL} {DODAG} version attacks},
  author={Mayzaud, Anth{\'e}a and Sehgal, Anuj and Badonnel, R{\'e}mi and Chrisment, Isabelle and Sch{\"o}nw{\"a}lder, J{\"u}rgen},
  booktitle={IFIP international conference on autonomous infrastructure, management and security},
  pages={92--104},
  year={2014},
  organization={Springer}
}
@inproceedings{mavani2017experimental,
  title={Experimental study of IP spoofing attack in 6LoWPAN network},
  author={Mavani, Monali and Asawa, Krishna},
  booktitle={Cloud Computing, Data Science \& Engineering-Confluence, 2017 7th International Conference on},
  pages={445--449},
  year={2017}
}
@inproceedings{varga2017security,
  title={Security threats and issues in automation IoT},
  author={Varga, Pal and Plosz, Sandor and Soos, Gabor and Hegedus, Csaba},
  booktitle={Factory Communication Systems (WFCS), 2017 IEEE 13th International Workshop on},
  pages={1--6},
  year={2017}
}
@article{singh2010hello,
  title={Hello flood attack and its countermeasures in wireless sensor networks},
  author={Singh, Virendra Pal and Jain, Sweta and Singhai, Jyoti},
  journal={IJCSI International Journal of Computer Science Issues},
  volume={7},
  number={11},
  pages={23--27},
  year={2010}
}
@article{garcia2013security,
  title={Security Considerations in the IP-based Internet of Things draft-garcia-core-security-06},
  author={Garcia-Morchon, O and Kumar, S and Keoh, S and Hummen, R and Struik, R},
  journal={Internet Engineering Task Force},
  year={2013}
}
@article{revathi2012survey,
  title={A survey of cooperative black and gray hole attack in MANET},
  author={Revathi, B and Geetha, D},
  journal={International Journal of Computer Science and Management Research},
  volume={1},
  number={2},
  pages={205--208},
  year={2012}
}
@inproceedings{pu2019energy,
  title={Energy Depletion Attack Against Routing Protocol in the Internet of Things},
  author={Pu, Cong},
  booktitle={16th IEEE Annual Consumer Communications \& Networking Conference (CCNC)},
  pages={1--4},
  year={2019},
  organization={IEEE}
}
@inproceedings{osterlind2006cross,
  title={Cross-level sensor network simulation with cooja},
  author={{\"O}sterlind, Fredrik and Dunkels, Adam and Eriksson, Joakim and Finne, Niclas and Voigt, Thiemo},
  booktitle={First IEEE International Workshop on Practical Issues in Building Sensor Network Applications (SenseApp 2006)},
  year={2006}
}
@article{DBLP:journals/twc/KhabbazianMB09,
  author    = {Majid Khabbazian and
               Hugues Mercier and
               Vijay K. Bhargava},
  title     = {Severity analysis and countermeasure for the wormhole attack in wireless
               ad hoc networks},
  journal   = {{IEEE} Trans. Wireless Communications},
  volume    = {8},
  number    = {2},
  pages     = {736--745},
  year      = {2009}
}
@inproceedings{dunkels2004contiki,
  title={Contiki-a lightweight and flexible operating system for tiny networked sensors},
  author={Dunkels, Adam and Gronvall, Bjorn and Voigt, Thiemo},
  booktitle={29th annual IEEE international conference on local computer networks},
  pages={455--462},
  year={2004},
  organization={}
}

@article{DAOinsider,
  title={Addressing the {DAO} Insider Attack in {RPL’s} Internet of Things Networks},
  author={Ghaleb, Baraq and Al-Dubai, Ahmed and Ekonomou, Elias and Qasem, Mamoun and Romdhani, Imed and Mackenzie, Lewis},
  journal={IEEE Communications Letters},
  volume={23},
  number={1},
  pages={68--71},
  year={2019},
  publisher={IEEE}
}

@inproceedings{le2013impacts,
  title={The impacts of internal threats towards routing protocol for low power and lossy network performance},
  author={Le, Anhtuan and Loo, Jonathan and Luo, Yuan and Lasebae, Aboubaker},
  booktitle={Computers and Communications (ISCC), IEEE Symposium on},
  pages={789--794},
  year={2013}
}
@article{karakehayov2005using,
  title={Using REWARD to detect team black-hole attacks in wireless sensor networks},
  author={Karakehayov, Zdravko},
  journal={Wksp. Real-World Wireless Sensor Networks},
  pages={20--21},
  year={2005}
}
@inproceedings{martin2004denial,
  title={Denial-of-service attacks on battery-powered mobile computers},
  author={Martin, Thomas and Hsiao, Michael and Ha, Dong and Krishnaswami, Jayan},
  booktitle={Proceedings of the Second IEEE Annual Conference on Pervasive Computing and Communications (PerCom), },
  pages={309--318},
  year={2004}
}
@inproceedings{stajano1999resurrecting,
  title={The resurrecting duckling},
  author={Stajano, Frank},
  booktitle={International workshop on security protocols},
  pages={183--194},
  year={1999}
}
@article{mosenia2017comprehensive,
  title={A comprehensive study of security of internet-of-things},
  author={Mosenia, Arsalan and Jha, Niraj K},
  journal={IEEE Transactions on Emerging Topics in Computing},
  volume={5},
  number={4},
  pages={586--602},
  year={2017}
}
@techreport{shelby2012neighbor,
  title={Neighbor discovery optimization for IPv6 over low-power wireless personal area networks (6LoWPANs)},
  author={Shelby, Zach and Chakrabarti, Samita and Nordmark, E and Bormann, C},
  year={2012}
}
@article{gomez2012problem,
  title={Problem statement and requirements for IPv6 over low-power wireless personal area network (6LoWPAN) routing},
  author={Gomez, Carles and Kim, Eunsook and Kaspar, Dominik and Bormann, Carsten},
  year={2012}
}
@article{kim2012design,
  title={Design and application spaces for IPv6 over low-power wireless personal area networks (6LoWPANs)},
  author={Kim, Eunsook and Kaspar, Dominik},
  year={2012}
}
@article{thubert2011compression,
  title={Compression format for IPv6 datagrams over IEEE 802.15.4-based networks},
  author={Thubert, Pascal and Hui, Jonathan W},
  year={2011}
}
@techreport{montenegro2007transmission,
  title={Transmission of IPv6 packets over IEEE 802.15. 4 networks},
  author={Montenegro, Gabriel and Kushalnagar, Nandakishore and Hui, Jonathan and Culler, David},
  journal= {RFC},
  volume= {4944},
  year={2007}
}
@techreport{kushalnagar2007ipv6,
  title={IPv6 over low-power wireless personal area networks (6LoWPANs): overview, assumptions, problem statement, and goals},
  author={Kushalnagar, Nandakishore and Montenegro, Gabriel and Schumacher, Christian},
  year={2007}
}
@article{hui2008extending,
  title={Extending IP to low-power, wireless personal area networks},
  author={Hui, Jonathan W and Culler, David E},
  journal={IEEE Internet Computing},
  volume={12},
  number={4},
  pages={37--45},
  year={2008},
  publisher={IEEE}
}
@article{palattella2013standardized,
  title={Standardized protocol stack for the internet of (important) things},
  author={Palattella, Maria Rita and Accettura, Nicola and Vilajosana, Xavier and Watteyne, Thomas and Grieco, Luigi Alfredo and Boggia, Gennaro and Dohler, Mischa},
  journal={IEEE communications surveys \& tutorials},
  volume={15},
  number={3},
  pages={1389--1406},
  year={2013},
  publisher={IEEE}
}
@article{granjal2015security,
  title={Security for the internet of things: a survey of existing protocols and open research issues},
  author={Granjal, Jorge and Monteiro, Edmundo and Silva, Jorge S{\'a}},
  journal={IEEE Communications Surveys \& Tutorials},
  volume={17},
  number={3},
  pages={1294--1312},
  year={2015},
  publisher={IEEE}
}
@article{perazzo2017dio,
  title={{DIO} Suppression Attack Against Routing in the Internet of Things},
  author={Perazzo, Pericle and Vallati, Carlo and Anastasi, Giuseppe and Dini, Gianluca},
  journal={IEEE Communications Letters},
  volume={21},
  number={11},
  pages={2524--2527},
  year={2017},
  publisher={IEEE}
}
@inproceedings{pongle2015survey,
  title={A survey: Attacks on RPL and 6LoWPAN in IoT},
  author={Pongle, Pavan and Chavan, Gurunath},
  booktitle={Pervasive Computing (ICPC), 2015 International Conference on},
  pages={1--6},
  year={2015}
}
@article{mayzaud2016taxonomy,
  title={A Taxonomy of Attacks in {RPL}-based Internet of Things},
  author={Mayzaud, Anth{\'e}a and Badonnel, R{\'e}mi and Chrisment, Isabelle},
  journal={International Journal of Network Security},
  volume={18},
  number={3},
  pages={459--473},
  year={2016}
}
@inproceedings{kyriazis2013sustainable,
  title={Sustainable smart city IoT applications: Heat and electricity management \& Eco-conscious cruise control for public transportation},
  author={Kyriazis, Dimosthenis and Varvarigou, Theodora and White, Daniel and Rossi, Andrea and Cooper, Joshua},
  booktitle={World of Wireless, Mobile and Multimedia Networks (WoWMoM), 2013 IEEE 14th International Symposium and Workshops on a},
  pages={1--5},
  year={2013}
}
@article{wood2002denial,
  title={Denial of service in sensor networks},
  author={Wood, Anthony D and Stankovic, John A},
  journal={computer},
  volume={35},
  number={10},
  pages={54--62},
  year={2002},
  publisher={IEEE}
}
@article{alexander2015security,
  title={A Security Threat Analysis for the Routing Protocol for Low-Power and Lossy Networks ({RPL}s)},
  author={Alexander, Roger and Tsao, Tzeta and Daza, Vanesa and Lozano, Angel and Richardson, Michael and Dohler, Mischa},
  year={2015}
}
@techreport{coltun2008ospf,
  title={{OSPF} for {IPv6}},
  author={Coltun, Rob and Ferguson, Dennis and Moy, John and Lindem, A},
  institution={A},
  year={2008}
}
@techreport{oran1990osi,
  title={{OSI} {IS-IS} intra-domain routing protocol},
  author={Oran, David},
  year={1990}
}
@techreport{perkins2003ad,
  title={Ad hoc on-demand distance vector ({AODV}) routing},
  author={Perkins, Charles and Belding-Royer, Elizabeth and Das, Samir},
  year={2003}
}

@inproceedings{landsmann2013topology,
  title={Topology authentication in {rpl}},
  author={Landsmann, Martin and Wahlisch, Matthias and Schmidt, Thomas C},
  booktitle={Computer Communications Workshops (INFOCOM WKSHPS), 2013 IEEE Conference on},
  pages={73--74},
  year={2013}
}
@inproceedings{dvir2011vera,
  title={{VeRA}-version number and rank authentication in {RPL}},
  author={Dvir, Amit and Buttyan, Levente and others},
  booktitle={Mobile Adhoc and Sensor Systems (MASS), IEEE 8th International Conference on},
  pages={709--714},
  year={2011}
}

@article{8270652, 
title={Compression Header Analyzer Intrusion Detection System (CHA -IDS) for 6LoWPAN Communication Protocol}, 
author={M. N. Napiah and M. Y. I. Idris and R. Ramli and I. Ahmedy}, 
journal={IEEE Access}, 
year={2018}, 
volume={PP}, 
number={99}, 
pages={1-1}, 
}
@inproceedings{hummen20136lowpan,
  title={6LoWPAN fragmentation attacks and mitigation mechanisms},
  author={Hummen, Ren{\'e} and Hiller, Jens and Wirtz, Hanno and Henze, Martin and Shafagh, Hossein and Wehrle, Klaus},
  booktitle={Proceedings of the sixth ACM conference on Security and privacy in wireless and mobile networks},
  pages={55--66},
  year={2013}
}
@article{ahmed2016mitigation,
  title={Mitigation of blackhole attacks in Routing Protocol for Low Power and Lossy Networks},
  author={Ahmed, Firoz and Ko, Young-Bae},
  journal={Security and Communication Networks},
  volume={9},
  number={18},
  pages={5143--5154},
  year={2016},
  publisher={Wiley Online Library}
}

@inproceedings{khan2013wormhole,
  title={Wormhole attack prevention mechanism for RPL based LLN network},
  author={Khan, Faraz Idris and Shon, Taeshik and Lee, Taekkyeun and Kim, Kihyung},
  booktitle={Ubiquitous and Future Networks (ICUFN), Fifth International Conference on},
  pages={149--154},
  year={2013}
}
@article{cragie2013applicability,
  title={Applicability Statement: The use of the {RPL} protocol set in Home Automation and Building Control},
  author={Cragie, Robert and Stok, Peter Van der and Brandt, Anders and Baccelli, Emmanuel},
  journal={Consultant},
  year={2013}
}

@article{raza2013svelte,
  title={SVELTE: Real-time intrusion detection in the Internet of Things},
  author={Raza, Shahid and Wallgren, Linus and Voigt, Thiemo},
  journal={Ad hoc networks},
  volume={11},
  number={8},
  pages={2661--2674},
  year={2013},
  publisher={Elsevier}
}
@techreport{rfc6550,
  title={{RPL}: {IPv6} routing protocol for low-power and lossy networks},
  author={ Tim Winter and et. al},
  institution={IETF},
  year={2012}
}
@article{RaoofML19,
  author    = {Ahmed Raoof and
               Ashraf Matrawy and
               Chung{-}Horng Lung},
  title     = {Routing Attacks and Mitigation Methods for {RPL}-Based Internet of Things},
  journal   = {{IEEE} Communications Surveys and Tutorials},
  volume    = {21},
  number    = {2},
  pages     = {1582--1606},
  year      = {2019},
  url       = {https://doi.org/10.1109/COMST.2018.2885894},
  doi       = {10.1109/COMST.2018.2885894},
  timestamp = {Fri, 05 Jul 2019 09:40:56 +0200},
  biburl    = {https://dblp.org/rec/bib/journals/comsur/RaoofML19},
  bibsource = {dblp computer science bibliography, https://dblp.org}
}
@inproceedings{weekly2012evaluating,
  title={Evaluating sinkhole defense techniques in {RPL} networks},
  author={Weekly, Kevin and Pister, Kristofer},
  booktitle={Network Protocols (ICNP), 20th IEEE International Conference on},
  pages={1--6},
  year={2012}
}

@article{wallgren2013routing,
  title={Routing Attacks and Countermeasures in the {RPL}-based Internet of Things},
  author={Wallgren, Linus and Raza, Shahid and Voigt, Thiemo},
  journal={International Journal of Distributed Sensor Networks},
  volume={9},
  number={8},
  pages={314--326},
  year={2013},
  publisher={SAGE Publications Sage UK: London, England}
}
@article{ahmad,
	author={Andrews, J.G. and Buzzi, S. and Wan Choi and Hanly, S.V. and Lozano, A and Soong, AC.K. and Zhang, J.C.},
	journal={IEEE Journal on Selected Areas in Communications},
	title={What Will 5{G} Be?},
	year={2014},
	month={June},
	volume={32},
	number={6},
	pages={1065-1082},
}